\documentclass[onecolumn, numberedappendix]{aastex6}
\usepackage{graphicx}
\usepackage{epsfig}
\usepackage{natbib}
\usepackage{amssymb}
\usepackage{amsbsy}
\usepackage{natbib}
\usepackage{url}
\usepackage[mathcal]{euscript}
\bibpunct{(}{)}{,}{a}{}{,}

\newcommand{\rsun}{R$_{\sun}$}
\newcommand{\msun}{M$_{\sun}$}

\newcommand{\eg}{{\it e.g.}}
\newcommand{\etal}{et~al.}

\begin{document}
 
\def\simlt{\vcenter{\hbox{$<$}\offinterlineskip\hbox{$\sim$}}}
\def\simgt{\vcenter{\hbox{$>$}\offinterlineskip\hbox{$\sim$}}}
%\simlt and \simgt produce > and < signs with twiddle underneath
\def\etal{et al.\ }
\def\kms{km s$^{-1}$}

\title{Pleiades or Not? Resolving the Status of the Lithium Rich M Dwarfs HHJ339 and HHJ430}
\author{John Stauffer\altaffilmark{1},
David Barrado\altaffilmark{2},
Trevor David\altaffilmark{3},
Luisa M. Rebull\altaffilmark{4},
Lynne A. Hillenbrand\altaffilmark{5},
Eric E. Mamajek\altaffilmark{6,7},
Rebecca Oppenheimer\altaffilmark{8},
Suzanne Aigrain\altaffilmark{9},
Herve Bouy\altaffilmark{10},
Jorge Lillo-Box\altaffilmark{11}
}
\altaffiltext{1}{Spitzer Science Center (SSC), IPAC, California Institute of
Technology, Pasadena, CA 91125, USA}
\altaffiltext{2}{Centro de Astrobiolog\'ia, Dpto. de
Astrof\'isica, INTA-CSIC, E-28692, ESAC Campus, Villanueva de
la Ca\~nada, Madrid, Spain}
\altaffiltext{3}{Center for Computational Astrophysics, Flatiron Institute, New York, NY 10010}
\altaffiltext{4}{Infrared Science Archive (IRSA), IPAC, California Institute of
Technology, 1200 E. California Blvd,
MS 100-22, Pasadena, CA 91125 USA}
\altaffiltext{5}{Astronomy Department,
California Institute of Technology, Pasadena, CA 91125 USA}
\altaffiltext{6}{Jet Propulsion Laboratory, California Institute of Technology, 4800 Oak Grove
  Drive, Pasadena, CA  91109, USA}
\altaffiltext{7}{Department of Physics \& Astronomy, University of Rochester, Rochester, NY 14627, USA}
\altaffiltext{8}{American Museum of Natural History , NYC, NY 12345}
\altaffiltext{9}{Sub-department of Astrophysics, Department of Physics,
University of Oxford, Oxford, OX1 3RH, UK}
\altaffiltext{10}{Laboratoire d{'}Astrophysique de Bordeaux, Univ. Bordeaux, CNRS, B18N, 
All\'ee Geoffroy Saint-Hillaire, F-33615 Pessac, France}
\altaffiltext{11}{Depto. de Astrof\'isica, Centro de Astrobiolog\'ia (CSIC-INTA), ESAC campus
28692 Villanueva de la Canada (Madrid), Spain}

\email{stauffer@ipac.caltech.edu}

\begin{abstract}

Oppenheimer \etal\ (1997) discovered two M5 dwarfs in the Pleiades with
nearly primordial lithium.   These stars are not low enough in mass to
represent the leading edge of the lithium depletion boundary at Pleiades
age ($\sim$125 Myr).  A possible explanation for the enhanced lithium 
in these stars is
that they are actually not members of the Pleiades but instead are members
of a younger moving group seen in projection towards the Pleiades.   We
have used data from Gaia DR2 to confirm that these two stars, HHJ 339 and
HHJ 430, are indeed not members of the Pleiades. Based on their space
motions, parallaxes and positions in a Gaia-based CMD, it is probable that
these two stars are about 40 parsecs foreground to the Pleiades and have
ages of $\sim$25 Myr.  Kinematically they are best matched to the 32 Ori
moving group.

\end{abstract}

\section{Introduction}

Star forming regions and young open clusters provide the laboratory data
for how star-formation and early stellar evolution proceed.  This only
works, however, if it is possible to attach ages to each of the laboratory
populations.   The more accurate the ages, the better the historical
reconstruction.   It was realized more than sixty years ago\footnote{Based
on spectra  obtained at the Crossley reflector by K. Hunger, while he was
visiting Lick Observatory and working with G. Herbig, as reported in the
1957 Annual Report of  Lick Observatory - Shane, C.D.  1957, AJ 62, 294.}
that the photospheric lithium abundance in low mass stars might provide one
means to determine those ages.  Very young low mass stars in star-forming
regions usually have nearly primordial lithium abundances (Bonsack 1959;
Bonsack \& Greenstein 1960).  There is a clear decrease in the mean lithium
abundance as a function of mass as one goes from stars of a few Myr age
(e.g. Orion or Taurus star-forming region, hereafter SFR) to stars of order
100 Myr (e.g. Pleiades) to stars of order 600 Myr (\eg\ Hyades) age
(Sestito, Palla \& Randich 2008; Soderblom \etal\ 1993; Cummings \etal\
2017).  While this dependence is clear when comparing data for large
ensembles of stars, there is significant dispersion in lithium abundance at
a given mass, such that it is not possible to assign accurate ages on a
star-by-star basis.

In the early 1990s, it was realized that lithium might become a quite
accurate age indicator for objects with masses near 0.1 \msun\ (Bildsten
1997), because below a certain mass the core temperature never becomes hot
enough to burn lithium, and these fully-convective objects should therefore
retain their primordial lithium abundance forever. Measuring the mass below
which all stars (and substellar objects) in a young open cluster still
retain nearly primordial lithium abundance therefore was  predicted to
provide a quite accurate age for all the stars in the cluster, assuming
that the stars in the cluster are essentially coeval.  The first cluster
for which an accurate ``lithium depletion boundary" (LDB) age was  measured
was the Pleiades (age 125 Myr, Stauffer \etal\ 1998). Subsequently, LDB
ages have been derived for the open clusters Alpha Persei, Blanco 1, NGC
1960, NGC 2516, NGC 2547, IC 2391, IC 4665, and Hyades  (Stauffer \etal\ 1999; 
Cargile, James and Jeffries 2010; Jeffries \etal\ 2013; Jeffries, James \&
Thurston 1998; Jeffries \& Oliveira 2005; Barrado, Stauffer \& Jayawardhana
2004;  Manzi \etal\ 2008; Martin \etal\ 2018) and for the Beta Pic and Tuc-Hor moving groups
(Binks \& Jeffries 2014; Kraus \etal\ 2014). 

In one of the earliest attempts to determine the lithium depletion boundary
in an open cluster, Oppenheimer \etal\ (1997) obtained spectra of a sample
of the faintest Pleiades members drawn from the Hambly, Hawkins \& Jameson
(1993; HHJ) proper motion survey.  They were unsuccessful in their quest
because the faint limit of the HHJ survey was just slightly brighter than
the location of the LDB in the Pleiades.  However, they did discover that
two of the moderately late (spectral type M5) cluster members (namely
HHJ 339 and HHJ 430) did have
strong lithium absorption features.  Because many fainter members did
not have lithium, those stars could not mark the location of the LDB in the
Pleiades unless there was a huge age spread in the cluster.  Oppenheimer
considered several possible explanations for the two stars with strong
lithium, but found none to be compelling.  The model with the fewest
problems was that the two stars were in fact not members of the Pleiades
but were instead members of a young moving group that happen to lie in our
line of sight to the Pleiades at the current time.  No subsequent paper has
attempted to more definitively  explain the abundant lithium in the spectra
of these two stars.

With the new evidence now available, we demonstrate that these two stars are indeed
foreground to the Pleiades and that their properties are most consistent
with membership in the 32 Ori moving group (Bell \etal\ 2017).  In \S 2, we
discuss the new data that we utilize in this paper.  In \S 3, we use Gaia
DR2 parallaxes and proper motions and our new radial velocities to show
that the two stars are  definitely not members of the Pleiades.   In \S 4,
we discuss the K2 light curves for the two stars, and argue that the light
curve for HHJ 339  suggests that it is younger than the Pleiades.   In \S5,
we show that HHJ 339 and 430 \footnote{In SIMBAD, these stars are referred
to as Cl* Melotte 22 HHJ 339, for example.} are likely members of the 32
Ori moving group based on their Gaia properties and the other data we 
present.

\section{Data Used in This Paper}

We use member lists for the 125 Myr old Pleiades cluster, the $\sim$ 25 Myr
old 32 Ori moving group, the Group 29 moving group (Oh \etal\ 2017; Luhman 2018), and
the $\sim$ 3 Myr old Taurus star-forming group in several of the plots we
will show.    These membership lists  are not intended as the complete set
of members, but are instead representative subsets of the members of those
groups (selected because they have particularly accurate radial velocities
in the literature or because they have particularly accurate astrometry).
The Pleiades list comes from the Gaia DR2 paper providing membership and HR
diagram morphologies for all the nearby open clusters (Gaia Collaboration,
Babusiaux, van Leeuwen \etal\ 2018a).  The radial velocities we use for the
Pleiades come from Mermilliod, Mayor, \& Udry (2009).  The member list for
Group 29 comes  from Oh \etal\ (2017); the member list for the 32 Ori group
comes from Bell, Murphy \& Mamajek 2017).  The Taurus member list is based
on  Rebull \etal\ (2020), which in turn heavily relies on the list from
Luhman (2018) and Esplin \& Luhman (2019); the Taurus radial velocities are
from Galli \etal\ (2019).

The Pleiades was observed by K2 (Howell \etal\ 2014) during Campaign 4.  
Processed light curves from that campaign were produced by several groups
(as described in Stumpe \etal\ 2012; Vanderburg \& Johnson 2014; Aigrain \etal\ 2016; Cody
\& Hillenbrand 2018).   Rebull \etal\ (2016a) used light curves from all of
those sources (selecting the best light curve for each star from among the
several choices) to determine rotation periods for all probable and
possible members of the Pleiades.   Light curves for both HHJ 339 and 430
were included in that analysis.  In Rebull \etal\ (2016a),  HHJ 430 was
ultimately considered to be a non-member of the Pleiades based on its
location in the CMD relative to true Pleiades members; HHJ 339 was
categorized as a possible but lower quality member (Bouy \etal\ 2015 reached
essentially the same conclusions regarding these two stars).
In \S 4, we provide a
detailed discussion of the K2 light curves of both stars. The relevance of
those light curves is primarily in that some light curve morphologies occur only
in young stars, and their presence (or absence) in the two HHJ 
stars could therefore help determine whether membership in the Pleiades is
likely or not.

We obtained new Keck HIRES spectra for both HHJ 339 and 430 in December
2013.   The spectra cover $\lambda\lambda$ 4800-9200\AA, at an average
resolution of about R=50,000, and typical S/N per pixel of about 30.  
A description of the data reduction
procedures and the process to determine radial velocities and $v \sin i$
values can be found in (David \etal\ 2019). From this analysis, we get RV =
11.3 $\pm$\ 5 \kms\ and $v \sin i$ = 45-55 \kms\ for HHJ 339, and 
RV = 15.8 $\pm$\ 5 \kms\
and $v \sin i$ = 50-55 \kms\ for HHJ 430. Oppenheimer \etal\ reported
slightly higher $v \sin i$ (58 and 65 \kms\ for HHJ 339 and 430,
respectively) and slightly lower RVs (9.4 and 9.1 \kms\ for HHJ 339 and
430, respectively), based on their HIRES spectra, with quoted uncertainties
of 5 \kms\ for each of the radial velocity and $v \sin i$ values. Figure 1
shows snippets from the two spectra centered on H$\alpha$\ and on the
\ion{Li}{1} $\lambda$6708 \AA\ region.  The \ion{Li}{1} equivalent widths
from our spectra (0.61 \AA\ for HHJ339 and 0.63 \AA\ for HHJ 430) are
consistent with  those reported by Oppenheimer et al.; the H$\alpha$
profiles and equivalent widths are consistent with those expected for young,
active, relatively late-type dMe stars.

High resolution images, taken with the lucky imaging technique,
were obtained with the Calar Alto 2.2m telescope and the Astralux
instrument during the night of 2015 November 20 in order
to obtain diffaction-limited images within the 24\arcsec $\times$ 24\arcsec 
FoV.  We used the AstraLux pipeline (see Hormuth \etal\ 2007) to
perform the basic reduction and combination of our lucky imaging
frames.  The Lucky imaging for both stars showed no evidence of any
companion, with a limit of about $\Delta m \sim$ 6 mag at 0.3 arcseconds
in each case.

\begin{figure}[ht]
\epsscale{0.9}
\plotone{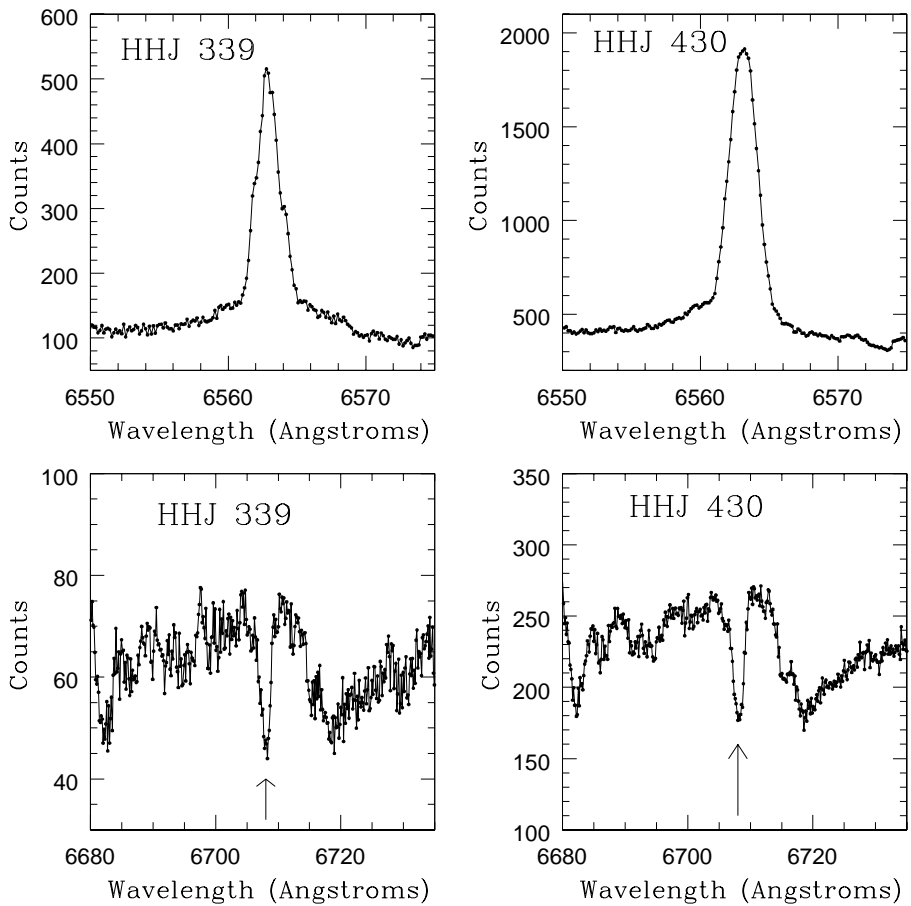}
\caption{(top) HIRES spectra showing the H$\alpha$\ emission profiles
for HHJ 339 and 430; (bottom) Keck HIRES spectra showing the  
\ion{Li}{1} 6708 spectral region for the two HHJ stars.  Arrow marks
the location of the lithium doublet.
\label{fig:Figure1}}
\end{figure}

The Gaia DR2 data release (Gaia Collaboration 2018b) provides by far the
most accurate parallaxes and proper motions for essentially all of the
stars we discuss in this paper. The Gaia photometry ($G$,
$B_p$ and $R_p$) for these stars is also the most accurate and homogeneous database from
which to construct a color-magnitude diagram for the cluster.  We have
downloaded the Gaia data from Vizier for all of the Pleiades members
identified in the DR2 HR diagram paper, as well as for the Taurus and young
moving group members we have investigated to help establish the true
lineage of HHJ 339 and 430.  In the next section, we use these data as the
primary evidence that the two HHJ stars are, in fact, not Pleiades
members.

\section{Implications from Gaia DR2 Data and The Measured Radial Velocities}

The Gaia DR2 data definitively resolved the Pleiades distance controversy
(van Leeuwen 2009; Abramson 2018), placing the Pleiades at a mean distance
of 135 pc (Lodieu \etal\ 2019), and not at the $\sim$120 pc distance that
had been inferred from Hipparcos data.  The Gaia release also provides the
best resource from which to  determine whether HHJ 339 and 430 are Pleiades
members or not.

Figure 2 shows the location of HHJ 339 and 430 in relation to the
known members of the Pleiades in RA/DEC space.   Both stars are seen in projection to
be relatively close to the center of the cluster.   The tidal radius of
the Pleiades has been estimated as $\sim$16 parsecs 
(Raboud \& Mermilliod 1998).   When projected onto the sky, that tidal
radius would lie entirely outside the region shown in Figure 2.  More
than 300 of the $\sim$1300 Gaia DR2 Pleiades members lie further
from the cluster center as projected on the sky than do HHJ 339 and 430.
Therefore, there is nothing in
the sky-projected spatial location of HHJ 339 and 430 
that argues against membership in the Pleiades.

\begin{figure}[ht]
\epsscale{0.9}
\plotone{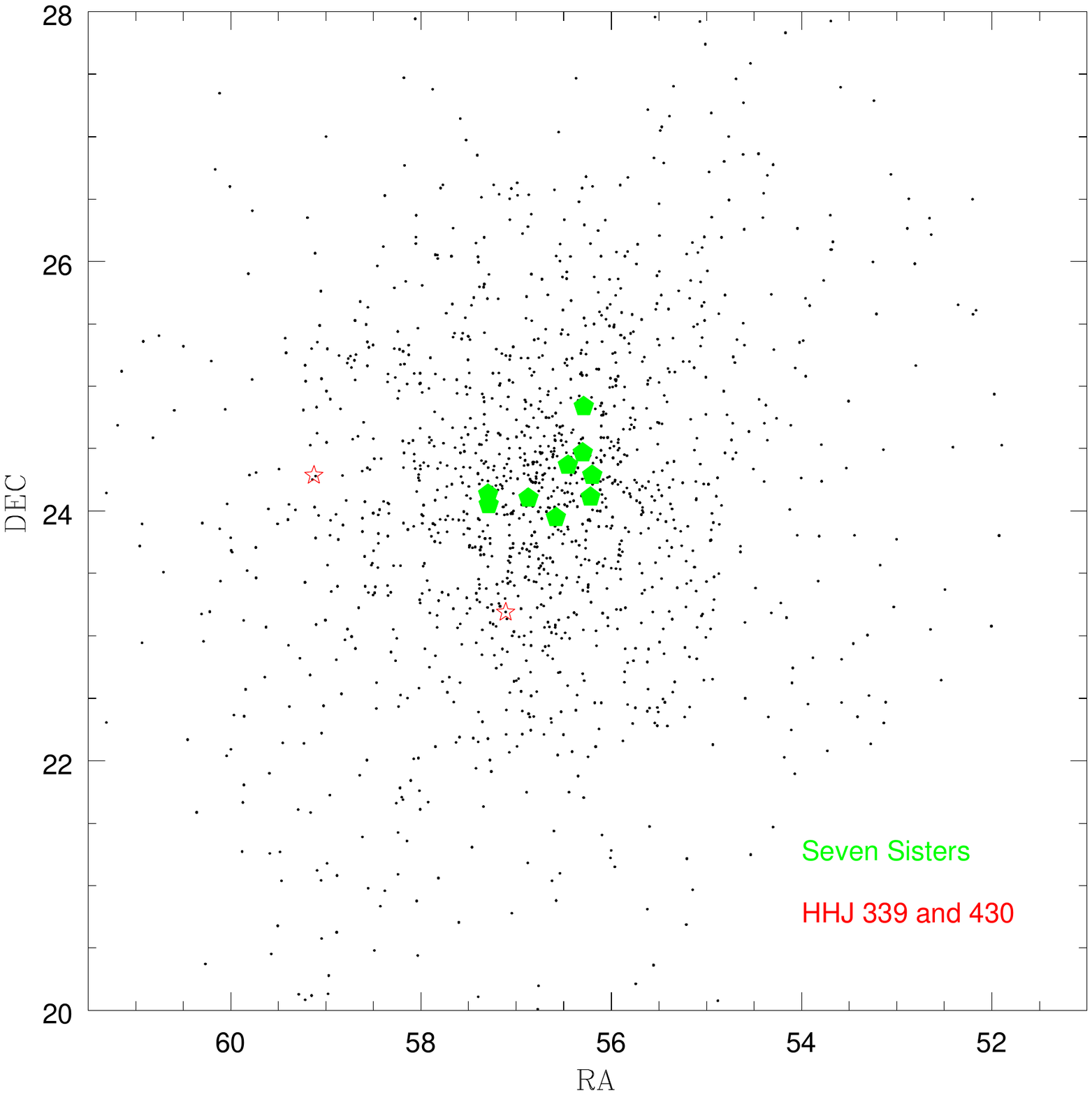}
\caption{Sky map of known Pleiades members.   The Seven Sisters (Alcyone,
Merope, Maia, Electra, Sterope, Taygete, and Celaeno) plus their parents
(Atlas and Pleione) are highlighted as large, filled circles.   HHJ 339 and
430 are shown as red stars.   Many low mass Pleiades members are located
outside the region plotted (the tidal radius of the Pleiades when projected
onto the sky corresponds to about 7 degrees).
\label{fig:figure2}}
\end{figure}

Figure 3a provides a visual comparison of the parallaxes for HHJ 339 and
430 relative to all of the high quality members of the Pleiades identified
using the DR2 data release.    With parallaxes larger than 10 mas, both HHJ
339 and 430 are much closer to us than the true Pleiades members.  The
median uncertainty in the parallax for the Pleiades members is 0.1 mas
yr$^{-1}$; for the two HHJ stars, the median parallax uncertainty
is a bit larger but still less than 0.2 mas yr$^{-1}$.  The two HHJ stars
are displaced to the foreground of the Pleiades by about 40 parsecs,
placing them well outside the tidal
radius of the cluster. 

Figure 3b shows a vector-point diagram (VPD) for the Pleiades members again
using the Gaia DR2 data, and again highlighting the positions of the two
lithium rich M dwarfs.   The true Pleiades members have proper
motions centered near 20 mas yr$^{-1}$ in RA and -45 mas yr$^{-1}$ in Dec. 
The HHJ stars have proper motions in right ascension that are about
10 mas yr$^{-1}$ larger than the mean Pleiades motion, much greater than
the $<$0.5 mas yr$^{-1}$ proper motion uncertainties typical of all the
stars plotted here.   At Pleiades distance, 10 mas yr$^{-1}$ corresponds to
about 7 \kms, which is much larger than the $\sim$0.8 \kms\ internal
velocity dispersion of the Pleiades (Galli \etal\ 2017).

\begin{figure*}[ht]
\epsscale{1.0}
\plotone{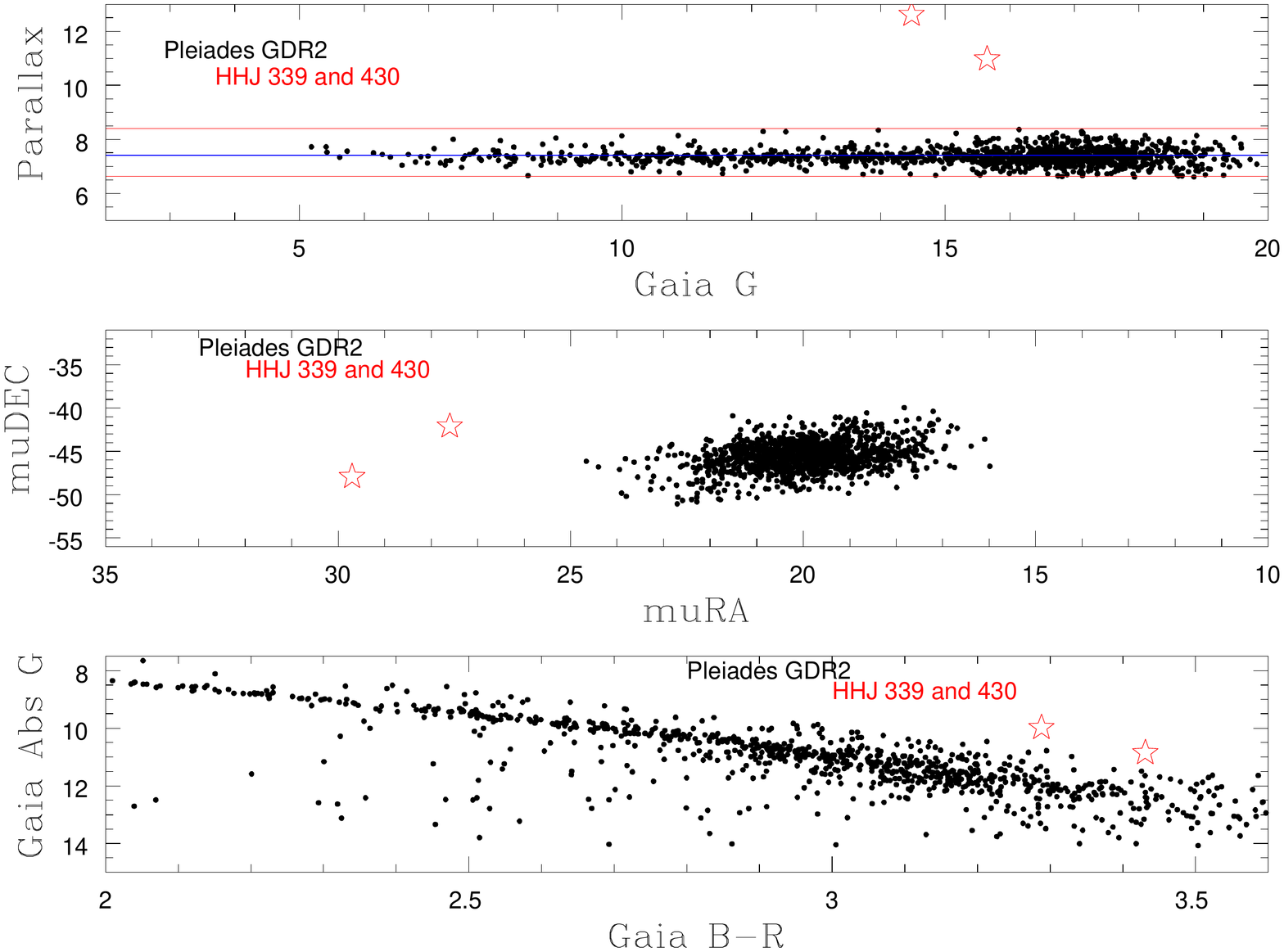}
\caption{(a)Gaia DR2 parallaxes of the Pleiades members  plotted versus
their Gaia G magnitude.  Red stars show the same data for the two
lithium rich M dwarfs HHJ 339 and 430.  The blue horizontal
line marks the mean parallax of the Pleiades; the two red horizontal lines
denote the tidal radius of the cluster.  (b) Gaia DR2 proper motions for
Pleiades members.   The two HHJ stars are again shown as red stars. (c)
Gaia-based CMD for the Pleiades members  plus the two lithium
rich M dwarfs.   In all three diagrams, the DR2 uncertainties in the plotted
quantities for HHJ 339 and 430 are much smaller than the size of the star
symbol used to mark their location.  The two HHJ stars are strong outliers in all three
diagrams and are clearly not Pleiades members.
\label{fig:superfigure3}}
\end{figure*}

Figure 3c shows a Gaia-based color-magnitude diagram (CMD) for the
Pleiades,  and the locations of HHJ 339 and 430 in that diagram.  Both of
the HHJ stars are  displaced above the single star locus by more
than 1.5 mag, hence above where even a triple system composed of equal mass
stars could be.  Both of the Oppenheimer stars must therefore be
significantly younger than 125 Myr.

Thus by every quantitative measure using the Gaia DR2 data,  HHJ 339 and
430 are demonstrably not Pleiades members.  Based on our own HIRES spectra
as well as that from Oppenheimer (1997), both stars do have nearly
primordial lithium, which for their Teff implies an age $<$ 40 Myr (Baraffe
\etal\ 2015; David \etal\ 2019).   We will attempt to better constrain
their ages after a brief digression concerning their photometric
variability.

\section{Kepler K2 Light Curves}

High precision, 70+ day light curves for both HHJ 339 and 430 were obtained
during Campaign 4 of NASA's K2 mission.  The K2 data for HHJ 430 shows two
strong periods, indicating that it is a binary star\footnote{G or K dwarfs can
have significant latitudinal differential rotation; their light curves can
exhibit two well-defined periods if they have spot groups located at widely
different latitudes.  Fully convective M dwarfs like HHJ 430 are expected
instead to have little or no latitudinal differential rotation, and therefore
two periods in their periodogram are best interpreted as evidence for the
presence of two stars in the system.   See Rebull \etal\ 2016b and
Stauffer \etal\ 2016 for further 
discussion of this point.}; the two periods are
0.3446 and 0.3736 days; such short periods would be fairly typical at
Pleiades age but atypically short at, for example, the $\sim$ 8 Myr age of
Upper Sco stars (see Rebull \etal\ 2018).  The two periods are quite similar
to each other, and the Lucky imaging shows that the two stars must also be
close to each other spatially.  The light curve morphologies
for both components of  HHJ 430 (shown in Figure 15 of Rebull \etal\
2016a)  are typical of that for rapidly rotating M dwarfs, where the
variability  is due to cool starspots.  However, this light curve
morphology puts little quantitative constraint on the age of HHJ 430.

By contrast, the K2 light curve for HHJ 339 shows a feature that is very
distinctive, and which has at least the potential to place a reasonably
quantitative constraint on its age.   Figure 4 shows the K2 light curve for
HHJ 339, phased to its period of 0.4627 day (Rebull \etal\ 2016a).  Based
on our visual examination of thousands of K2 light curves, the entire shape
of this light curve seems unusual, possibly pointing to something other
than non-axisymmetrically distributed spots as the physical  mechanism
responsible for the photometric variability.  However, it is possible that
some unusual distribution of spots could more or less explain most of the
variability shown in Figure 4.   What spots cannot explain, however, is the
relatively deep and narrow in phase flux dip centered near phase 0.75.  As
argued in a number of papers (Stauffer \etal\ 2017; David \etal\ 2017; Zhan
\etal\ 2019), flux dips such as this are most likely due to dust
``clouds" orbiting at the Keplerian corotation radius that  pass through
our line of sight to the star.  The variability of the shape of the dip on
timescales less than a K2 campaign length ($\sim$ 75 days) -- see Figure 4
-- is typical of some of these stars, including RIK-210 (David \etal\ 2017)
and a few of the other PMS M dwarfs in Upper Sco (Stauffer \etal\ 2017,
2018).   Such narrow-in-phase flux dips are very rare or absent at ages
older than the Pleiades (Rebull \etal\ 2018; Basri \& Nguyen 2018).   With
existing data, it is not yet possible to place a quantitative age
constraint on HHJ 339 based on the presence, depth and shape of its narrow
flux dip, but by combining data from K2, TESS and Gaia for open clusters
and moving groups of a variety of ages, such a quantitative age constraint
may become possible.

By combining the $v \sin i$, rotation period, and the position of HHJ 339
in an HR diagram, we can estimate the inclination angle of the star's
rotational axis.  To convert the photometry and spectral type information
for HHJ 339 into luminosity and effective temperature, we adopt the data
tables in Pecaut \& Mamajek (2013, hereafter PM13).   In order to use those
tables, however, we need a rough estimate of the true age of HHJ 339.   In
the next section, we will adopt an age of 25 Myr for HHJ 339, which implies
we should use Table 6 of PM13 to provide the correlation between $V-K_s$
color or spectral type and $T_{\rm eff}$, and the bolometric correction
appropriate for that $T_{\rm eff}$.  Photometry from 2MASS for HHJ 339
includes $V$=17.45 (Kamai \etal\ 2014), $J$=12.164 and $K_s$=11.32
(Skrutskie \etal\ 2006).  Adopting $A_V$\ = 0.12, this yields $(V-K_s)_o$ =
6.02, from which Table 6 of PM13 yields $T_{\rm eff}$ = 2900K and BC$_J$ =
2.00.   Combining those numbers with the Stefan-Boltzmann equation then
yields $R$ = 0.48 \rsun. Combining this with the measured period and the
average of our $v \sin i$ estimate and Oppenheimer's, we derive $\sin i$ =
1.0, and hence that the system inclination is near 90 degrees.  This is
roughly as expected for a model where the occulting material is in a plane
located between the equatorial rotation plane and the equatorial plane of
the star's dipole magnetic field (Jardine \etal\ 2020).   

\begin{figure}[ht]
\epsscale{0.9}
\plotone{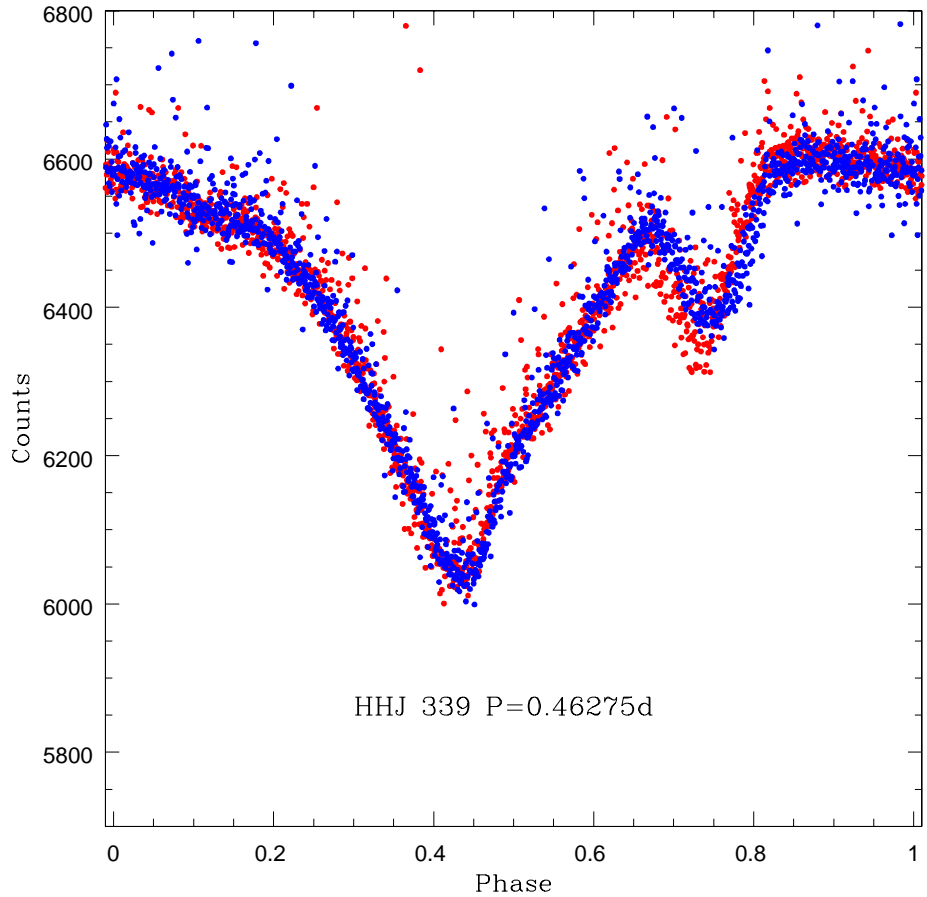}
\caption{K2 light curve for HHJ 339, phased to its rotation period of
P=0.4627 days.  Blue points indicate data in the first half of the K2
campaign period; red points indicate data taken during the second half of
the campaign.  The shape of the narrow flux dip at phase $\sim$ 0.75
changes between the two time periods, whereas the rest of the light curve
morphology remains nearly  constant over the whole K2 campaign.  Narrow
flux dips like this are only present in the optical light curves of young,
low mass stars. \label{fig:superfigure4}}
\end{figure}

\section{Age and True Lineage of HHJ 339 and HHJ 430}

The preceding section provides strong evidence that HHJ 339 and 430 are 
not members of the Pleiades, and are in fact foreground to the Pleiades and
much younger.  Can we  accurately determine the age of these two stars and
learn something of their true origin?  

There is a long and at times contentious history concerning the age spread
within the Taurus SFR and/or the presence and extent of a relatively young
moving group population towards the general direction of Taurus.   Wichmann
\etal\ (1996) identified a set of more spatially extended, apparently
slightly older low mass PMS stars in the general direction of Taurus and
posited that they were real members of the Taurus SFR and therefore
evidence for a significant age spread.   Briceno \etal\ (1997)
instead argued that the older, more extended population of stars were
members of one or more moving groups seen in projection toward the Taurus
SFR but not natally connected to it.   Dozens of papers have been published
arguing this issue since the 1990s.  The Gaia DR2 data offer the
possibility to at least largely settle the issue (Luhman 2018; Kraus \etal\
2019).  Based on a lengthy analysis of the DR2 data and other published
sources, both Luhman and Kraus \etal\ concluded that the more spatially
extended population most probably represents a previous generation of star
formation, unconnected to the $\sim$3 Myr old Taurus SFR population.   
Group 29 (Oh \etal\ 2017) and the 32 Ori Group (Bell, Murphy \& Mamajek
2017) have space motions, ages and spatial distributions that make them
likely contributors to the older-but-still-young, spatially extended
population of stars toward Taurus. We compare the properties of HHJ 339 and
430 to the members of Group 29, the 32 Ori Group and Taurus in the
following plots.

Figure 5 shows space motions of the two Oppenheimer lithium rich M dwarfs
compared to that of members of the other young stellar groups that populate
our line of sight toward Taurus.   We have used the on-line website 
(http://kinematics.bdnyc.org/query) to convert measured proper motions,
radial velocities, positions and distances to $UVW$ space motions (Rodriguez 2016); for the
radial velocities of the HHJ stars, we adopt the average of our values and those of
Oppenheimer \etal\ 1997.   The
space velocity plots show that when accurate input data are used, the space
motions of HHJ 339 and 430 are  inconsistent with Pleiades membership and
also with membership in Group 29. Their space motions are most consistent
with that of the 32 Ori group.

\begin{figure}[ht]
\epsscale{0.9}
\plotone{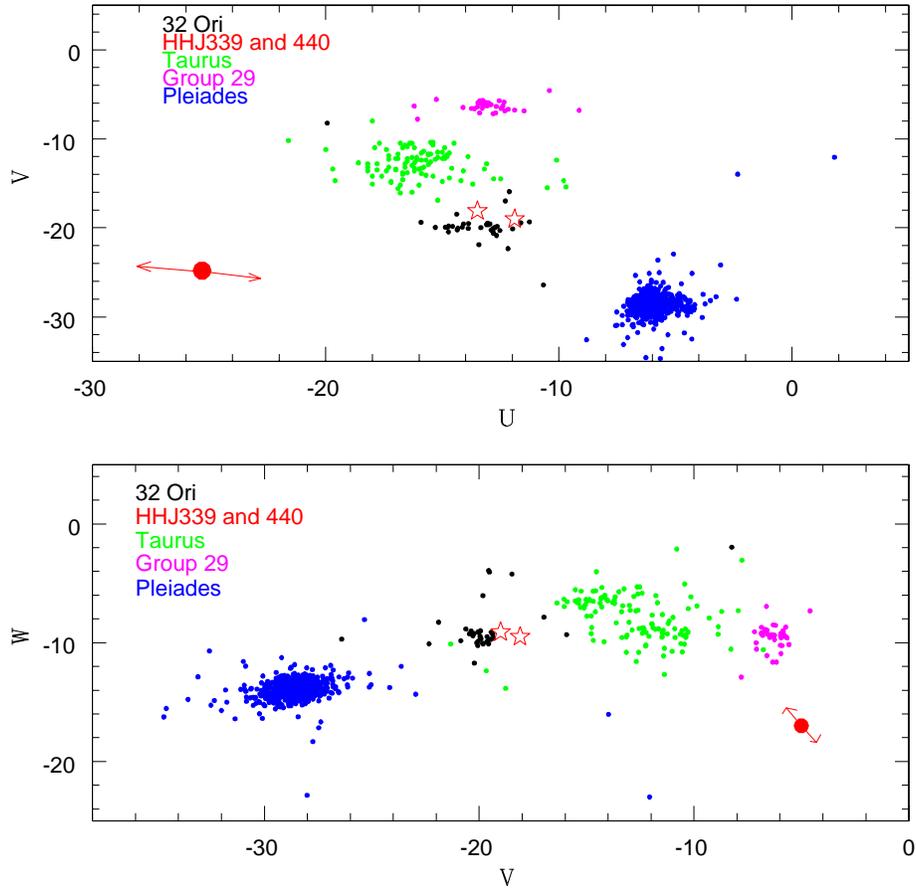}
\caption{Space motion plots for the lithium richh M dwarfs HHJ 339 and 430
compared to other kinematic groups known to be present in the general 
direction of K2 fields 4 and 13.  By far the largest source of uncertainty
in the UVW motions of HHJ 339 and 430 are $\sim$3.5 \kms\ uncertainties in their
radial velocities.  The red dot and associated arrows in each figure shows
the impact of the one sigma uncertainty in radial velocity on their derived
UVW motions.  The two HHJ  stars have kinematics
that are quite disparate from the Pleiades, but are most compatible with
the 32 Ori moving group.
\label{fig:superfigure5}}
\end{figure}

In Figure 3c, we showed that HHJ 339 and 430 had locations in a Gaia-based
CMD that were incompatible with membership in the Pleiades (they are both
too bright for their $B_p-R_p$ color).  Figure 6 shows another CMD, this
time plotting probable members of the 32 Ori and Group 29 moving groups 
along with the two HHJ lithium rich M dwarfs.  The two moving
groups appear to have quite similar isochronal ages; both moving groups
have estimated ages of about 25 Myr (David \etal\ 2019; Bell \etal\ 2017).
The two Oppenheimer stars have locations in this CMD consistent also with
that age.

\begin{figure}[ht]
\epsscale{0.9}
\plotone{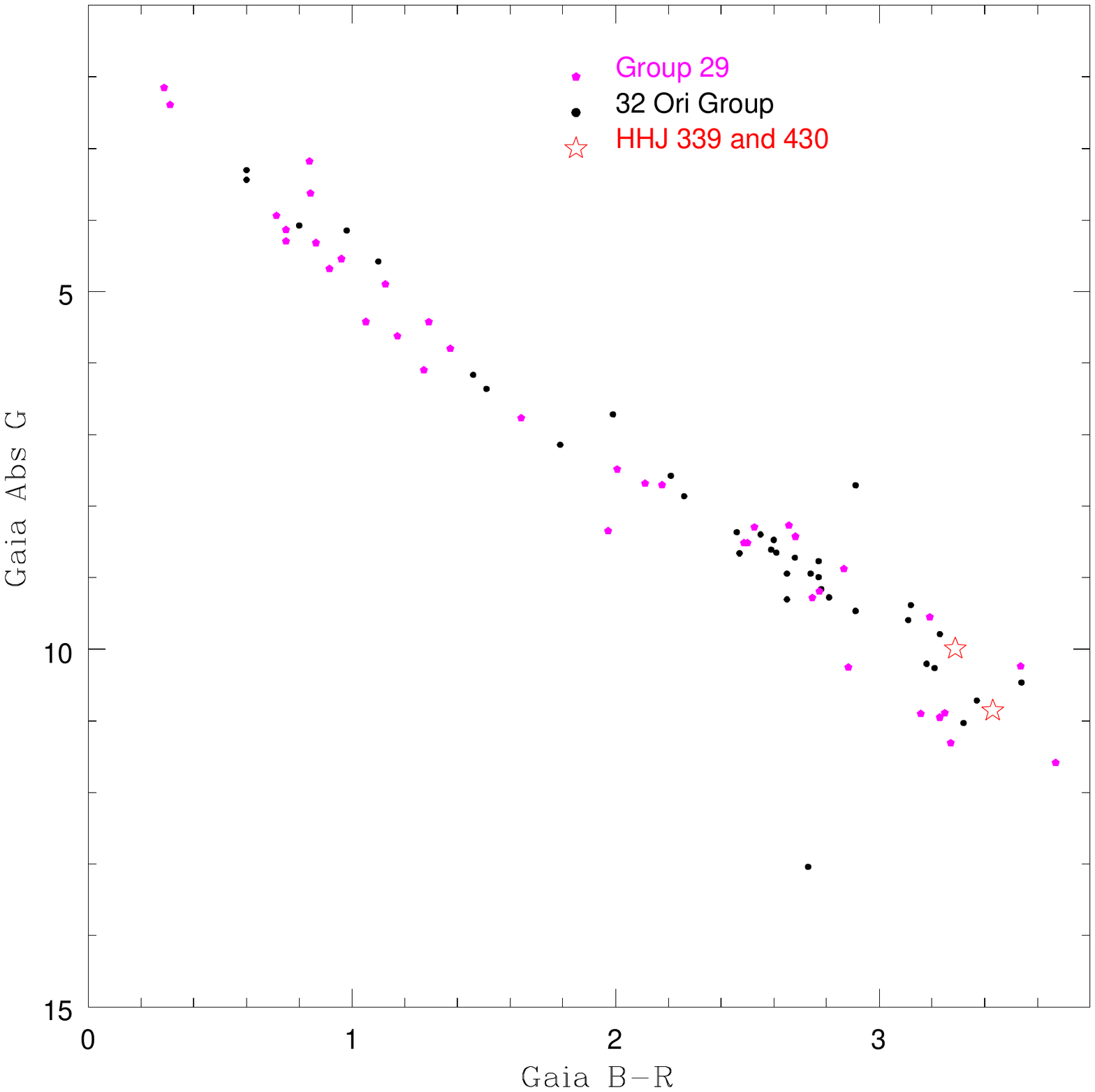}
\caption{Gaia-based CMD for the members of the 32 Ori group and the two
Oppenheimer lithium rich M dwarfs.   The Oppenheimer stars have CMD
locations consistent with the 32 Ori group members.
\label{fig:superfigure6}}
\end{figure}

Table 1 shows the mean space motions of the systems we have discussed, as
well as that for the Beta Pic moving group.  The table again shows that the
space motions of HHJ 339 and 430 are best aligned with that for the 32 Ori
moving group.   However, given the $\sim$ 3.5 \kms\ uncertainty in the radial
velocity for the HHJ stars, membership in 
the Beta Pic moving group (BPMG) cannot be excluded.  The age estimated for the BPMG
ranges from 10-30 Myr, but has recently been reported to be near 20-25 Myr
(Binks \& Jeffries 2014; Mamajek \& Bell 2014; Bell \etal\ 2015),  and so is quite similar to
that for the 32 Ori moving group.

\floattable
\begin{deluxetable*}{lccc}
\tabletypesize{\footnotesize}
%\rotate
\tablecolumns{4}
\tablewidth{0pt}
\tablecaption{Space Motions of Systems Relevant to this Paper\label{tab:basicdata}}
\tablehead{
\colhead{ ID} &
\colhead{U} &
\colhead{V} &
\colhead{W}  \\
\colhead{} &
\colhead{\kms} &
\colhead{\kms} &
\colhead{\kms}  }
\startdata
Pleiades   & -6.2 & -28.7 & -14.7  \\
Taurus SFR  & -14.3 &  -9.3 & -8.8  \\
32 Ori group  & -12.8 & -18.8 & -9.9  \\
Group 29  & -13. & -6. &  -9.5  \\
\hline \\
HHJ 339  & -11.8 & -19.0 & -9.1  \\
HHJ 430  & -13.5 & -18.1 & -9.5  \\
\hline  \\
Beta Pic MG  & -10.9 & -16.0 & -9.0  \\ 
\enddata
\end{deluxetable*}
\noindent

\section{Discussion and Conclusions}

The Pleiades is the most intensively studied open cluster in the sky.  Its
membership list is correspondingly quite heavily vetted, with very few
stars whose membership were greatly in debate even prior to the advent of
space-based astrometric missions.  HHJ 339 and 430 were exceptions to that
rule.  They had been identified as probable Pleiades members based on their
proper motions.   They had independently been identified as young dM stars
seen in the direction of the Pleiades based on their being flare stars and
on their being strong X-ray sources, criteria that in most cases
successfully selects Pleiades members.   The fact that their spectra show a
nearly primordial lithium abundance therefore came as a surprise.   If they
were indeed members of the Pleiades, then some exotic physics must be
involved (late accretion of large, rocky bodies?) or the Pleiades must
contain an admixture of stars much younger than the main population.   If
they are not members of the Pleiades, then there must be a previously
unsuspected population of young stars projected onto the face of the
Pleiades.   

Publication of the Gaia DR2 catalog has provided the resolution to this
conundrum. The proper motions of HHJ 339 and 430, while similar to that of
Pleiades members, are inconsistent with Pleiades membership when measured
to the accuracy provided by the DR2 data.  Both stars are also
significantly foreground to the Pleiades based on the exquisite DR2
parallaxes.   When combined with radial velocities from Keck HIRES spectra,
we find that HHJ 339 and 430 have space motions that match that of  the 32
Ori moving group.   They also have photometry that matches that of
previously identified 32 Ori members when a Gaia-based CMD is
constructed.   At the estimated age of 25 Myr ascribed to the 32 Ori group,
models predict that stars with M5 spectral type should retain nearly
primordial lithium abundance, thereby explaining the original anomaly
discovered by Oppenheimer \etal

We thank all those who helped build and operate the Gaia satellite and
those who worked hard to analyse the data and produce the astrometric and
photometric catalogs that are now available.   This paper could not have
been written without their labor.

\begin{acknowledgements}

Some of the data presented in this paper were obtained from the Mikulski
Archive for Space Telescopes (MAST). Support for MAST for non-HST data is
provided by the NASA Office of Space Science via grant NNX09AF08G and by
other grants and contracts. This paper includes data collected by the
Kepler mission. Funding for the Kepler mission is provided by the NASA
Science Mission directorate. This research has made use of the NASA/IPAC
Infrared Science Archive (IRSA), which is operated by the Jet Propulsion
Laboratory, California Institute of Technology, under contract with the
National Aeronautics and Space Administration. This research has made use
of data products from the Two Micron All-Sky Survey (2MASS), which is a
joint project of the University of Massachusetts and the Infrared
Processing and Analysis Center, funded by the National Aeronautics and
Space Administration and the National Science Foundation. The 2MASS data
are served by the NASA/IPAC Infrared Science Archive, which is operated by
the Jet Propulsion Laboratory, California Institute of Technology, under
contract with the National Aeronautics and Space Administration.  This
research has made use of NASA's Astrophysics Data System (ADS) Abstract
Service, and of the SIMBAD database, operated at CDS, Strasbourg, France. 
Part of this research was carried out at the Jet Propulsion Laboratory, California 
Institute of Technology, under a contract with the National Aeronautics and Space Administration.
DB and JLB have been funded by the Spanish State Research Agency (AEI) Projects
No. ESP2017-87676-C5-1-R and No. MDM-2017-0737 Unidad de Excelencia
``Mar\'ia de Maeztu" - Centro de Astrobiolog\'ia (INTA-CSIC).
\end{acknowledgements}

\facility{K2} \facility{Exoplanet Archive} \facility{IRSA}
\facility{2mass}

\end{document}